\documentclass[reprint,showpacs,preprintnumbers,prl,10pt,aps,groupedaddress]{revtex4-1}
\usepackage[pdftex]{graphicx}
\usepackage{dcolumn}
\usepackage{bm}
\usepackage{epsfig}
\usepackage{latexsym}
\usepackage{amsmath}
\usepackage{color}
\usepackage{array}

\newcommand{\<}{\langle}
\newcommand{\e}{\varepsilon}
\newcommand{\up}{\uparrow}
\newcommand{\down}{\downarrow}

\renewcommand{\>}{\rangle}
\renewcommand{\(}{\left(}
\renewcommand{\)}{\right)}
\renewcommand{\[}{\left[}
\renewcommand{\]}{\right]}

\begin{document}
\title{Superconductivity and Ferromagnetism in Oxide Interface Structures: Possibility of Finite Momentum Pairing}
\author{Karen Michaeli}
\affiliation{Department of Physics, Massachusetts Institute of Technology,
77 Massachusetts Avenue, Cambridge, MA 02139}
\author{Andrew C. Potter}
\affiliation{Department of Physics, Massachusetts Institute of Technology,
77 Massachusetts Avenue, Cambridge, MA 02139}
\author{Patrick A Lee}
\affiliation{Department of Physics, Massachusetts Institute of Technology,
77 Massachusetts Avenue, Cambridge, MA 02139}
\begin{abstract}
We introduce a model to explain the observed ferromagnetism and superconductivity in LAO/STO oxide interface structures.  Due to the polar catastrophe mechanism, 1/2 charge per unit cell is transferred to the interface layer.  We argue that this charge localizes and orders ferromagnetically via exchange with the conduction electrons.  Ordinarily this ferromagnetism would destroy superconductivity, but due to strong spin-orbit coupling near the interface, the magnetism and superconductivity can coexist by forming an FFLO-type condensate of Cooper pairs at finite momentum, which is surprisingly robust in the presence of strong disorder.
\end{abstract}

\maketitle

\emph{Introduction -- }It is known that a conducting electronic state can form at the interface between two insulating oxides\cite{Ohtomo04}. A particularly well studied example is the TiO$_2$ interface between SrTiO$_3$ and LaAlO$_3$.  The carrier density can be controlled by a backgate on the SrTiO$_3$ side and superconductivity (SC) has been discovered over a range of densities with maximum T$_c$ of about 0.3~K\cite{Reyren07}.  Recently, signs of ferromagnetism (FM) have also been reported\cite{Brinkman07,Dikin,Li,Bert11,Ariando11}. In particular, Li {\em et al}.\cite{Li} showed that SC and FM coexist in the same sample and that the FM moment is large, $\approx 0.3-0.4 \mu_B$ per interface unit cell.  Assuming that the FM and SC arise from the interface, these observations raise the question of whether the SC has to be unconventional in order to coexist with FM.  Before addressing this question we need to  understand first the nature of the electronic state at the interface and up to now no clear picture has emerged\cite{Okamoto06, Pentcheva07/08,Popovic08}. Are most of the electrons localized or extended? Does the FM come from local moments or the mobile electrons and what is its origin?  In this paper we propose a model for the interface electrons which is consistent with existing transport data.  Based on this model we explain the existence of FM and the coexistence of SC and FM.  For the latter, the key idea is that a large Rashba-type spin-orbit coupling exists at the interface \cite{Caviglia10}. Such a Rashba term is particularly favorable for the formation of a condensate at finite momentum, called a  Fulde-Ferrell-Larkin-Ovchinikov (FFLO) state which coexists with FM\cite{Fulde64,Larkin65}.   This general idea was pointed out earlier by Barzykin and Gorkov\cite{Barzykin02}.  However, they considered only the clean case and their solution is quickly destroyed by disorder.  Surprisingly, with increasing disorder the FFLO state is revived\cite{Dimitrova}.  We suggest that the SC observed at the interface is described by this disordered stabilized helical FFLO state.  This state is sometimes referred to as a ``helical" FFLO state\cite{Dimitrova} since, pairing occurs at a single momentum $\mathbf{q}$ so that $\Delta(\mathbf{r}) = \Delta e^{i\mathbf{q}\cdot\mathbf{r}}$ unlike the usual FFLO state where pairing occurs at both $\pm\mathbf{q}$ so that $\Delta(\mathbf{r}) = \Delta\cos\mathbf{q}\cdot\mathbf{r}$.

\emph{The Model -- }   As shown in Fig. 1, LaAlO$_3$ consists of layers with alternating charged, while SrTiO$_3$ has charged-neutral layers. As a result of the charge discontinuity at the interface, an electric potential  proportional to the number of LaAlO$_3$ layers  is built up. This phenomenon is termed the polar catastrophe. Since the Ti ions allow for mixed valence charge compensation, to avoid the polar catastrophe, half an electron per unit cell is transferred from the surface AlO$_2$ layer to the TiO$_2$ across the interface.  The electrons are expected to occupy the $d_{xy}$ orbital on the Ti atoms.  Due to the relatively narrow bandwidth, an on-site repulsion $U$ and a nearest-neighbor Coulomb repulsion $V$ will cause these electrons to be localized on every other interface site. This picture of local moment formation at the interface has been proposed before~\cite{Pentcheva07/08}.   Super-exchange via the oxygen is expected to provide a weak antiferromagnetic exchange.  

\begin{figure}[ttt]
\begin{center}
\includegraphics[width=3.3in]{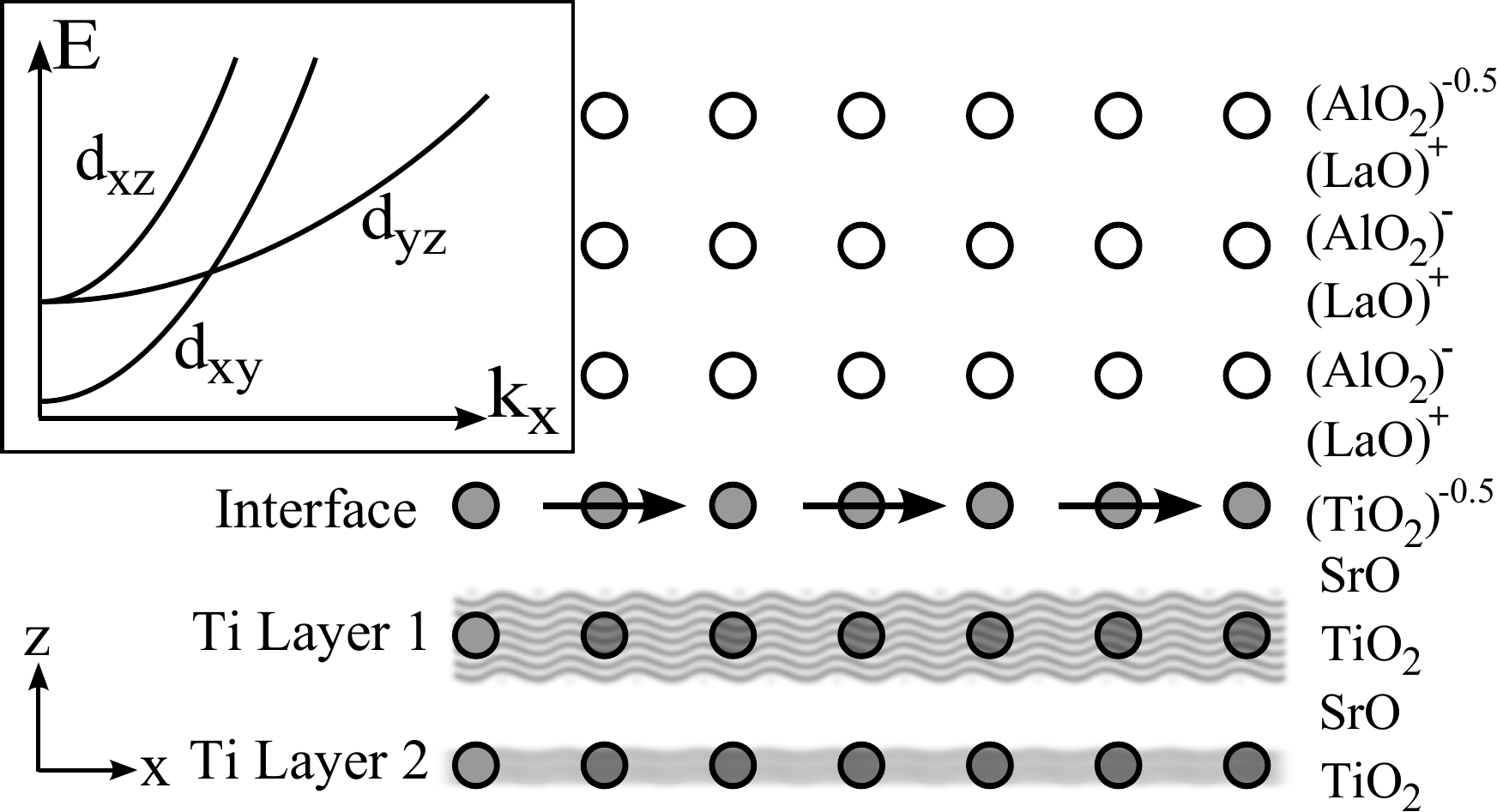}
\end{center}
\vspace{-.2in}
\caption{Schematic depiction of 
the SrTiO$_3$/LaAlO$_3$ 
oxide interface structure.  
Filled and empty circles depict Ti and Al ions.
The half electron charge per unit cell is transferred to the interface $\text{TiO}_2$ layer  localizes and orders magnetically (shown as arrows on the interface layer) via exchange polarization of conduction electrons on the subsequent Ti layers (shown as wavy cloud in Ti Layers 1 and 2).  Inset: Dispersion of electron bands arising from 3d orbitals on Ti layers near the interface.}
\label{fig:Interface}
\vspace{-.1in}
\end{figure}

The application of a back gate or the presence of defects  forces more electrons to the interface.  We assume that the effect of $U$ and $V$ makes it too costly to place these electrons at the interface layer.  Instead, the additional electrons mainly occupy the Ti layer next to it (called layer 1), and their wavefunctions  spill over to layers 2 and 3 as well.  We model these electrons as occupying two dimensional conducting bands.  These are the electrons seen in transport measurements, with a typical areal density of $3 \times 10^{13}$cm$^{-2}$, or about 10\% of the density compared with the localized electrons.    These electrons initially go into the $d_{xy}$ band~\cite{Joshua11}.  Hall effect measurements show nonlinearity in magnetic field $H$ with increasing gate voltage, which has been interpreted as the appearance of a second carrier with lower mobility\cite{BenShalom10,Bell09,Joshua11}.  We assume that these are the $d_{xz}$ and $d_{yz}$ bands, illustrated in the inset of  Fig.~1. These bands are highly anisotropic with a heavy mass in one direction which may be responsible for the lower mobility.  The $d_{xz}$ and $d_{yz}$ bands are higher in energy because their lobes point towards the negative charge at the interface and because their bandwidths are narrower.

In addition, transport measurements show that the elastic scattering rate $1/{\tau}$ drops rapidly with increasing carrier density~\cite{Bell09}.  Furthermore, from the analysis of magnetoresistance, it was found that a Rashba term $H_{\rm R} = \alpha \hat{z}\cdot(\boldsymbol{\sigma} \times \bm{k}) $ grows rapidly with the gate voltage~\cite{Caviglia10}, reaching a spin splitting $\Delta_{so} = 2\alpha k_F$ of 10~meV near the peak of the superconducting dome, a value comparable to the Fermi energy of $\approx 40$~meV.  Since a back gate voltage tends to pull carriers from the interface, the decrease of $1/{\tau}$ is reasonable but the increase of $\Delta_{so}$ is counterintuitive.  We believe this trend is a consequence of increasing admixture of the $d_{xz}$ and $d_{yz}$ bands with increasing carrier density.  The Rashba energy is determined by the polarization of the electron wave-function due to the asymmetric environment at the interface and the contributions come mainly from near the atomic core, where the electron is subject to a large electric field ${-\partial V}/{\partial z}$:
\begin{equation}
\Delta_{so} \propto |k_\parallel| \frac{2}{c^2} \int d{\bm r} \frac{dV}{dz} |\psi_{\bm k_\parallel} ({\bm r})|^2
\end{equation}
where the wavefunction $\psi_{\bm k_\parallel}({\bm r}) = \sum_\ell a_\ell \phi_{\bm k_\parallel}^\ell({\bm r})$
and $\ell$ denote various angular momenta which are admixed due to the asymmetric environment of the interface~\cite{Nagano09}.  Let us restrict ourselves to admixtures between $d$, $p$ and $s$ states.  Since $\nabla V \propto \hat{z}$, non-vanishing contributions involving the $d$ bands in Eq.(1) come only from the cross term between $d_{xz}$ and $p_{x}$ and between $d_{yz}$ and $p_y$.  Furthermore, the $d_{xy}$ band can have a nonzero $\Delta_{so}$ only via the admixture of 
either $s$ and $p$ or, more importantly of $d_{xz}$ with $p_y$ and $d_{yz}$ with $p_x$. We expect that the latter hybridization between $d$ and $p$ orbitals gives rise to $\Delta_{so}$ that increases with chemical potential.

The final ingredient of our model is the exchange coupling between the local moments and the conduction electrons.  We write the standard phenomenological form
\begin{equation}
H_J = J_K \sum_i \int d{\bm r} {\hat{\bm S}_i}\cdot {\hat{\bm s}}({\bm r})\delta({\bm R}_i - {\bm r})\end{equation} 
where
$
{\hat{\bm s}}(r) = \frac{1}{2}\psi_\alpha^\dagger {\bm\sigma}_{\alpha\beta}\psi_\beta(r)
$
is the electron spin density operator in the $d_{xy}$ band, and ${\hat{\bm S}}_i$ is the local spin operator on site $i$.  We introduce a similar coupling $J_K^\prime$ for the $d_{xz}$ and $d_{yz}$ bands.  It is useful to introduce $J_0 = J_K/n_0$ where $n_0$ is the inverse of the interface unit cell area, and similarly $J_0^\prime = J_K^\prime/n_0$.  The Schrieffer-Wolff expression is $J_0 = 2 {\tilde{t}}^2 (\frac{1}{U+ \varepsilon_d} - \frac{1}{\varepsilon_d})$ where ${\tilde{t}}$ is the hybridization between the local moment and the conduction band orbital and $\varepsilon_d < 0$ is the orbital energy of the local moment relative to the chemical potential. We note that the $d_{xz}$ and $d_{yz}$ orbitals in layer 1 are orthogonal to the localized $d_{xy}$ orbital in the interface layer, so that the hybridization ${\tilde{t}}^\prime$ vanishes except for the admixture of other orbitals in the $d_{xz}$ and $d_{yz}$ bands.  We therefore expect $J_0^\prime \ll J_0$.

\emph{The Origin of Ferromagnetism -- } We note that the present problem is the opposite limit to the familiar problem of dilute Kondo impurities, where the Kondo screening of the local moments competes with RKKY interactions between them.  Here the density of local moments $n_i = \frac{1}{2} n_0$ is much greater than the conduction electron density $n_c$, i.e., the separation between local moments $1/\sqrt{n_i}$ is much smaller than $k_F^{-1}$.  We can still view the local moments as interacting via RKKY interactions, but this interaction will be ferromagnetic and with a relatively long range of $(2k_F)^{-1}$.  The FM ordering temperature $T_F$ can be worked out \cite{Dietl97} and apart from a numerical constant, the result 
was shown to be 
equivalent to a mean field treatment of $H_J$ which we shall adopt below.  In this picture, which was introduced by Zener \cite{Zener51} and referred to as the Zener kinetic exchange mechanism, the local moments order by polarizing the conduction electrons.  This mechanism has been applied successfully to explain the FM of Mn substitution in GaAs and we borrow the results here~\cite{Jungwirth06}.  We introduce the average localized spin order per site ${\boldsymbol{\cal S}} = 
\frac{1}{N_i} \sum_i\<{\hat{\bm S}}_i\>$ and 
the average electron density ${\mathbf{ s}} = \frac{1}{\rm vol} \int d{\bm r}\<{\hat{\bm s}}({\bm r})\>$.  To quadratic order the total free energy density takes the form
\begin{equation}
E_{\rm tot} = \frac{1}{2} \frac{| \mu_0{\boldsymbol{\cal S}}|^2 }{\chi_0}n_i + 
\frac{1}{2} \frac{| \mu_0{\mathbf{s}}|^2 }{\chi_c} +
J_0 \frac{n_i}{n_0} {\boldsymbol{\cal S}} \cdot {\mathbf{s}} .
\end{equation}
The last term is the mean field decoupling of Eq.(2).  In the first term $\chi_0 = \frac{1}{3}\frac{\mu_0^2 S(S+1)}{T + \theta}$ where $\mu_0 = g\mu_B$ and $S = {1}/{2}$, $g = 2$ in what follows, $\theta > 0$ is the Weiss term due to the weak AF super-exchange exchange which we shall ignore below.  In the second term $\chi_c = \frac{1}{4} \mu_0^2 \nu(0)$ where $\nu(0) = m^\ast/\pi\hbar^2$ is the density of states including spin of a free electron gas.  
(The presence of a Rashba term does not change the spin susceptibility of a free electron gas~\cite{Barzykin02,Yip02}.)
Minimizing Eq.(3) with respect to ${\boldsymbol{\cal S}}$ leads to a purely quadratic term in $|{\boldsymbol{\cal S}}|^2$, and the sign change of its coefficient determines the FM transition temperature 
\begin{equation}
k_B T_F = \frac{S(S+1)}{12} J_0^2 \frac{n_i}{n_0} \frac{\nu(0)}{n_0} .
\end{equation}
We find that $\nu(0)/n_0 = 0.64 (m^\ast_{xy}/m_\text{e})$~eV$^{-1}$.  For $m^\ast_{xy}/m = 0.7$, $J_0 = 1.3$~eV will give the observed $T_F \approx 300$~K.  
By comparison, for Mn/GaAs, $J_0$ is $\approx 1$~eV. Here ${\tilde{t}}$ is smaller, but $|\varepsilon_d|$ is also smaller because the same orbital is involved in the local moment and the conduction electron, so the estimated $J_0$ appears reasonable.
Thus we conclude that the Zener kinetic exchange mechanism can account for a robust FM state.

Next we estimate the polarization of the conduction electron.  In the mean field theory, the effect of ${\boldsymbol{\cal S}}$ on the conduction electrons is described by an effective Zeeman field $
H_J \approx \int d{\bm{r}}\mu_0 \bm{H}_{\rm{MF}} \cdot \bm{s}(\bm{r})$ where ${\bm{H}}_{{\rm MF}}= J_0 \frac{n_i}{n_0}\boldsymbol{\cal S}$.
Li {\em et al}.\cite{Li} reported an ordered moment of $= 0.3 \mu_B$ per interface unit cell, i.e., $0.6 \mu_B$ per local moment in our picture, which implies $|\boldsymbol{\cal S} | = 0.3$.  Using $J_0 = 1.3$~eV, we estimate a Zeeman spin splitting $|\mu_0 {\bm H}_{\rm MF}| \approx 200$~meV, which is comparable to or exceeds the Fermi energy.  Thus the $d_{xy}$ band is largely spin polarized.  Since $J_0^\prime \ll J_0$, we expect a smaller (but still significant) polarization of the $d_{xz}$ and $d_{yz}$ bands.  

\emph{Nature of the Superconducting State -- } We assume that the superconductivity originates from a conventional electron-phonon coupling mechanism, which is modeled by an attractive short range interaction $g$ with a cut-off given by the Debye frequency $\omega_D$.  Since the superconducting transition temperature is T$_c \approx 0.3$~K, the pairing gap $\approx 0.04$meV is the lowest energy scale in the problem. In the $d_{xy}$ band we estimate a Zeeman splitting of 0.2~eV, which exceeds the Pauli limit by more than 3 orders of magnitude, and precludes the possibility of pairing in the $d_{xy}$ band.  The $d_{xz}$ and $d_{yz}$ bands will also be partially polarized due to the exchange interaction.  However, the exchange splitting, $\mu_0B$, in these bands is expected to be much smaller (although likely still $\mu_0B\gg\Delta$).  Moreover, as we argue above, we expect that the Rashba spin-orbit coupling, $\Delta_{so}$, is even larger in the $d_{xz}$ and $d_{yz}$ bands than that observed in the $d_{xy}$ bands.  It is natural to look to strong spin-orbit coupling to preserve pairing in the $d_{xz}$ and $d_{yz}$ bands despite large Zeeman splitting.

The enhancement of $B_c$ due to Rashba spin-orbit coupling was first demonstrated by \cite{Barzykin02} the case of weak Rashba coupling ($\Delta_{so}\ll \e_F$) and no disorder, and later by \cite{Dimitrova} for the case of weak Rashba coupling and moderate disorder.  They showed that an FFLO state is favored, where the pairing occurs with a finite center of mass  momentum\cite{GSCurrentEndNote}.  Here we extend their analysis to treat arbitrarily strong values of $\Delta_{so}$ and disorder.  We begin by neglecting disorder and find the susceptibility to form Cooper pairs at finite pair momentum $\mathbf{q}=q\hat{y}$.  The dispersion for the $\pm$ Rashba branches is: $\e^\pm_{{\bf k}+\mathbf{q}/2}= \frac{({\bf k}+\mathbf{q}/2)^2}{2m}-\mu\pm\alpha\sqrt{k_x^2+(k_y+q/2+\mu_0B/{\alpha})^2}$.\cite{AnisotropicDispersionSupplement}
In the physically relevant limit: $v_Fq,B\ll\Delta_{so}$, and we can expand in $q$ and $B$:
\begin{equation}\label{eq:dispersion}
 \e^\pm_{{\bf k}+\mathbf{q}/2}\approx \e^\pm_k(B=0)+(v_Fq/2\pm \mu_0B)\sin\phi_k +\mathcal{O}(v_Fq^2,B^2)
 \end{equation}
where $v_F = \sqrt{\alpha^2+2\mu/m}$ is the Fermi-velocity for the Rashba bands and $\phi_k = \tan^{-1}(\frac{k_y}{k_x})$.  The key is that choosing $q=\frac{2\mu_0B}{v_F}$\cite{SignAlphaEndnote} ensures $\e^{-}_{\mathbf{k}+\mathbf{q}/2}=\e^{-}_{-\mathbf{k}+\mathbf{q}/2}+\mathcal{O}(\frac{B^3}{\Delta_{so}^2})$ for \emph{all} angles $\phi_k$.  This should be contrasted with the usual FFLO case without spin-orbit coupling, where the linear terms cannot be cancelled for all angles for any choice of $q$.  However, we cannot prevent an energy mismatch in both bands simultaneously. By choosing ${\bf q}={2\mu_0B\hat{y}}/{v_F}$ we optimize for the $\e_-$ branch, which has a larger density of states, $\nu_-$, and find:
 \begin{equation} \mu_0B_c \approx \Delta_0\(\Delta_{so}/\Delta_0\)^{\frac{1+\alpha/v_F}{2+\alpha/v_F}}
\label{eq:BCEnhancementClean} \end{equation}
where $\Delta_0 = \omega_D\exp\[\frac{-1}{(\nu_++\nu_-)g}\]$ is the superconducting gap in the absence of the Zeeman field $\mu_0B$.  In the limit  $\alpha\ll{v_F}$  we recover the results of Barzykin and Gor'kov: $\mu_0B_c \approx {\Delta_0}\sqrt{{\Delta_{so}}/{\Delta_0}}$.  For the oxide interface system we expect stronger spin-orbit coupling, $\Delta_{so}\approx{\e_F}$, and find an even larger enhancement: $\mu_0B_c \approx {\Delta_0}(\Delta_{so}/{\Delta_0})^{2/3}$.  

\begin{figure}[ttt]
\begin{center}
\includegraphics[width=3in]{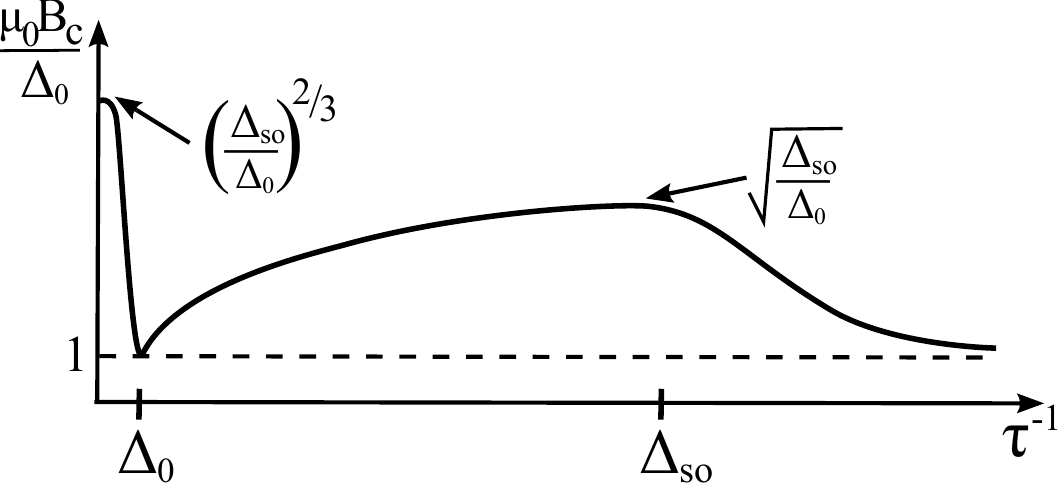}
\end{center}
\vspace{-.2in}
\caption{Critical Zeeman splitting, $B_c$, measured with respect to the bare SC gap $\Delta_0$ as a function of disorder strength $\tau^{-1}$. The FFLO state identified in \cite{Barzykin02} is rapidly destroyed as $\Delta_0\tau\rightarrow 1$.  Remarkably, for stronger disorder the FFLO state re-emerges and $B_c$ is enhanced beyond the Pauli limit (shown as a dashed line).}
\label{fig:BcDisorder}
\vspace{-.1in}
\end{figure}

The above calculation is only valid in very clean systems for which $\Delta_0\tau\gg 1$.   In practice, we expect to be in the dirty limit, $\Delta_0\tau\ll 1$.  To incorporate impurity scattering, we consider spin-less, short-ranged impurities and compute the disorder averaged Cooper-channel susceptibility in the $\e_F\tau\gg 1$ limit, by summing the ladder diagrams for impurity scattering (called the Cooperon).
As shown in Fig.~\ref{fig:BcDisorder}, there are 3 regimes. First, in the weak disorder regime ($\tau^{-1}<\Delta_0$) we find that the critical field drops rapidly to the Pauli limit $B_c\approx\Delta_0(\Delta_0\tau)^{\frac{1+\alpha/v_F}{1-\alpha/v_F}}$ . This can be understood as follows: as in the clean case, the pair momentum minimizes the effect of the magnetic field in the $\e_{-}$ branch, while keeping the pair breaking in the $\e_{+}$ branch. However, impurities can scatter  Cooper pairs from the $-$ band to the $+$ band where they rapidly decohere. Thus, disorder enhances the dephasing effects of the Zeeman field, which becomes fully pair breaking as $\tau^{-1}\rightarrow \Delta_0$. On the other hand, for very strong disorder, $\tau^{-1}\gg\Delta_{so}$, the Rashba bands $\e^\pm$ lose their identifies due to the rapid impurity scattering.  Here, spin and momentum become decoupled and the problem reduces to that of conventional parabolic bands with effective spin-orbit scattering rate $\tilde{\tau}_{so}^{-1}=\Delta_{so}^2\tau\ll\tau^{-1}$.  This is the D'yakonov-Perel limit where the spin diffuses in small steps between rapid impurity scattering\cite{Dyakonov}.  In this limit, it was demonstrated in \cite{Klemm} that $\mu_0B_c\approx \Delta_0/\sqrt{\Delta_0\tilde{\tau}_{so}}$, and SC occurs at $q=0$.

The interesting limit is the intermediate regime $\Delta_0\ll\tau^{-1}\ll\Delta_{so}$\cite{Dimitrova}.  Here the disorder is weak enough that the Rashba bands maintain their identity, but a pair in the $\e^+$ band can readily be scattered to a pair in the $\e^-$ band.  Unlike the weak disorder case, the pairing is strongly admixed and it is not possible to sacrifice the pair breaking of one band in favor of the other.  On the other hand,  due to the spin-orbit coupling both spin (the Zeeman energy) and momentum show diffusive behavior. As a consequence, the pair breaking effect of the magnetic field is weaker.  The pairing interaction is dominated by the Cooperon, $\mathcal{C}$, which ordinarily develops a diffusion pole $\mathcal{C}= \frac{1}{i\omega+Dq^2}$ with $D = \frac{v_F^2\tau}{2}$, but in the present case $q$ is replaced by $q\pm\frac{2\mu_0B}{v_F}$ for the $\pm$ bands respectively. Due to the strong mixing of the pairing channels, the effective pair breaking strength is given by the combination: $\sum_{\lambda=\pm} \frac{\nu_\lambda}{4m}\(v_Fq+\lambda 2\mu_0B\)^2$.  Since $\nu_+\neq\nu_-$ this combination is minimized by finite momentum $q=\frac{4\alpha B}{\alpha^2+v_F^2}$, and we predict that an FFLO state exists in the intermediate regime, but with a different $q$ from the clean case.  The corresponding $B_c$ is: $\mu_0B_c \approx \frac{\Delta_0 }{\sqrt{\Delta_0\tau}}$. Physically, on this limit $\tau_B\ll \tau$ and, unlike the weak disorder, the pair dephasing time $\tau_B^2/\tau$ grows with increasing disorder.

\emph{Discussion --} To summarize, we propose a model to explain the coexistence of SC and FM observed in STO/ALO oxide interface structures.  In this model, a half-charge per unit cell is transferred to the interface layer and forms a lattice of local moments due to Coulomb repulsion.  These local moments then order ferromagnetically via exchange with lower-density bands of mobile electrons residing in Ti layers near the interface.  The large FM exchange would ordinarily kill SC in these mobile bands.  However, the presence of a large spin-orbit coupling  enables the formation of an FFLO state which can coexist with strong magnetism.  In this FFLO state, Cooper pairs form with finite pair momentum perpendicular to the direction of magnetic ordering.  Unlike the usual FFLO state without spin-orbit coupling~\cite{Fulde64,Larkin65}, spin-orbit coupling  parametrically enhances $B_c$ beyond the Pauli limit and enables FFLO pairing to survive to much stronger disorder (up to $\Delta_{so}\tau\gtrsim 1$).  Experimentally, $\tau^{-1}$ decreases with back-gate voltage $V_G$ while simultaneously $\Delta_{so}$ increases.  In our picture, the observed $T_c$ vs $V_G$ dome is related to the behavior of $B_c$ shown in Fig. \ref{fig:BcDisorder} sweeping from $\tau^{-1}> \Delta_{so}$ to $\tau^{-1}<\Delta_{so}$.

We wish to emphasize that our model suggests that  the FM and SC occur in different bands, and are uniformly distributed at the interface. Another possible explanation of the coexistence is in the spatial separation of the two phases. Indeed, a recent experiment~\cite{Bert11} shows that  the FM is arrange in domains and might not occupy the entire interface layer. However, the large total magnetic moment of the system~\cite{Li}, implies that the domains should occupy most of the area at the interface. The SC order parameter, in contrast, does not go to zero at any point in the plane.


Before concluding, we briefly discuss some experimental signatures of our model.  First, if the exchange coupling for the $d_{xz}$ and $d_{yz}$ bands is such that the conduction electrons polarize in the opposite direction of the interface moments, then the external in-plane field $H_\parallel$ would align the local moments along $H_\parallel$ but would reduce the total Zeeman field seen by the$d_{xz}$ and $d_{yz}$ bands to $\mu_0(B-H_\parallel)$.  This leads to the unusual prediction that T$_c$ should exhibit a maximum at finite $H_\parallel$.  Second, one can look for $\mathbf{q}\neq 0$ pairing by creating a Josephson tunneling junction with a conventional SC film.  By applying a magnetic field parallel to the junction, Cooper pairs tunnel at finite momentum $\Delta \mathbf{k}$, and the Josephson current would peak when $\Delta \mathbf{k} = \mathbf{q}$. 

\emph{Acknowledgments --} We thank Lu Li and Ray Ashoori for many helpful discussions, and acknowledge support by the Pappalardo Fellowship at MIT (KM),  NSF IGERT Grant No. DGE-0801525 (ACP), and DOE grant FG02-03ER46076 (PAL).

\onecolumngrid
\vspace{.5in}
\begin{center} {\bf Supplementary Materials}\end{center}

In the main text we have explained that the conduction electrons at the interface can either occupy the $d_{xy}$ band or the $d_{xz}$ and $d_{yz}$ bands. According to our model, the electrons in the $d_{xy}$ band are polarized by the localized electrons and participate in establishing the the Ferromagnetic order. The exchange interaction between the electrons in the  $d_{xz}$ and $d_{yz}$ bands and both the localized electrons and those occupying the $d_{xy}$ band is much weaker. Then, these electrons can undergo a phase transition into the superconducting state. Still, the exchange field plays an important role as it create an effective Zeeman  field parallel to the interface, which is expected to exceed the Pauli limit.   As we argued before, the strong spin orbit interaction induced by the asymmetric environment at the interface, can stabilize the superconducting state at magnetic fields much higher than the  Pauli limit. This is done by pairing electrons with finite momentum (the FFLO state).

In this supplement we present a sketch of the derivation of the gap equation from which we determine the critical exchange field for the FFLO state.  In addition to the effects of spin-orbit coupling and the exchange field, we wish to examine the effect of disorder on the state.  The full details of the derivation and its extension to various special cases will be presented in a longer companion paper~\cite{CompanionPaperSupp}.

We start from the following Hamiltonian describing the conduction electrons in the $d_{xz} $ and $d_{yz}$ bands:
\begin{align}\label{eq:Hamiltonian}
H=H_{0}+H_{\text{imp}}+H_{\text{J}}+H_{\text{sc}},
\end{align}
where $H_{0}$ is the Hamiltonian of the electrons in the presence of Rashba spin-orbit interaction:
\begin{align}\label{eq:Hamiltonian0}
H_{0}=\sum_{\mathbf{k},\alpha}\frac{k^2}{2m}c_{\alpha}^{\dag}(\mathbf{k})c_{\alpha}(\mathbf{k})+\alpha\hat{z}\cdot\left(\boldsymbol{\sigma}_{\alpha\beta}\times\mathbf{k}\right)c_{\alpha}^{\dag}(\mathbf{k})c_{\beta}(\mathbf{k}).
\end{align}
The operator $c_{\alpha}^{\dag}(\mathbf{k})$ ($c_{\alpha}(\mathbf{k})$) is the creation (annihilation) of an electron with spin $\alpha$ along the $z-direction$ and momentum $\mathbf{k}$. Here we assume that the kinetic energy of the electrons can be approximated by a parabolic band.  In reality, the $d_{xz}$ and $d_{yz}$ bands will have different mass in the $x$ and $y$ directions, however, this does not qualitatively change the subsequent analysis (as we demonstrate below). We consider a random impurity potential:
\begin{align}\label{eq:Hamiltonian_imp}
H_{\text{imp}}=\sum_{r,\alpha}V(r)c_{\alpha}^{\dag}(r)c_{\alpha}(r).
\end{align}  
which has only short-range correlations: $\overline{V(r)V(r')} = V^2\delta(r-r')$.
The effective Zeeman field $\mathbf{B}\equiv\mathbf{H}_{MF}$ is chosen to be parallel to the $x$-axis. Thus, 
\begin{align}\label{eq:Hamiltonian_J}
H_{\text{J}}=\sum_{\mathbf{k},\alpha,\beta}\mu_{0}\mathbf{B}\cdot\boldsymbol{\sigma}_{\alpha\beta}c_{\alpha}^{\dag}(\mathbf{k})c_{\beta}(\mathbf{k}').
\end{align}  
Finally, the pairing  interaction is:
\begin{align}\label{eq:Hamiltonian_sc}
H_{\text{sc}}=-g\sum_{\mathbf{k},\mathbf{k}',\mathbf{q},\alpha,\beta}c_{\alpha}^{\dag}(\mathbf{q}/2+\mathbf{k})c_{-\alpha}^{\dag}(\mathbf{q}/2-\mathbf{k})c_{\beta}(\mathbf{q}/2+\mathbf{k}')c_{-\beta}(\mathbf{q}/2-\mathbf{k}').
\end{align}  

For the next stages of the derivation it will be helpful to transform into the chiral basis that diagonalizes $H_0+H_{J}$:
\begin{align}\label{eq:Chiral_Basis}
&c_{\up}(\mathbf{k})=\frac{1}{\sqrt{2}}\left[\psi_{+}(\mathbf{k})+\psi_{-}(\mathbf{k})\right];\\\nonumber
&c_{\down}(\mathbf{k})=\frac{k_{x}+i(k_{y}+\mu_0B/\alpha)}{\sqrt{2}|\mathbf{k}+\mu_0B\hat{y}/\alpha|}\left[\psi_{+}(\mathbf{k})-\psi_{-}(\mathbf{k})\right].
\end{align}  
where $\psi_\pm^\dagger$ ($\psi_\pm$) are electron creation (annihilation) operators in the chiral basis.
Correspondingly, the eigenvalues of electrons in the two chiral bands (also referred to as the Rashba branches) are:
 \begin{align}\label{eq:Chiral_Basis_Energy}
E_{\lambda}(\mathbf{k})=\frac{k^2}{2m}+\lambda\alpha\sqrt{k_x^2+(k_y+\mu_0B/\alpha)},
\end{align} 
 with $\lambda=\pm1$. In the chiral basis, both the impurity scattering matrix element as well as the superconducting interaction acquire momentum dependence:
\begin{subequations}\label{eq:Chiral_Basis_V+g}
\begin{align}\label{eq:Chiral_Basis_V}
&V_{\lambda,\lambda'}(\mathbf{k},\mathbf{k'})=\frac{V^2}{4}\left[1+\lambda\lambda'\frac{k_x-i(k_y+\mu_0B/\alpha)^2}{|\mathbf{k}+\mu_0B\hat{y}/\alpha|}
\frac{k_x+i(k_y+\mu_0B/\alpha)^2}{|\mathbf{k}+\mu_0B\hat{y}/\alpha|}\right];
\end{align}
\begin{align}\label{eq:Chiral_Basis_int}
H_{\text{sc}}=
-\sum_{\mathbf{k},\mathbf{k}',\mathbf{q}}g_{\lambda_1,\lambda_2,\lambda_3,\lambda_4}(\mathbf{k},\mathbf{k}',\mathbf{q})\psi_{\lambda_1}^{\dag}(\mathbf{q}/2+\mathbf{k})\psi_{\lambda_2}^{\dag}(\mathbf{q}/2-\mathbf{k})\psi_{\lambda_3}(\mathbf{q}/2+\mathbf{k}')\psi_{\lambda_4}(\mathbf{q}/2-\mathbf{k}'),
\end{align}  
where repeated indices are implicitly summed and,
\begin{align}\label{eq:Chiral_Basis_g}\nonumber
&g_{\lambda_1,\lambda_2,\lambda_3,\lambda_4}(\mathbf{k},\mathbf{k}',\mathbf{q})=\frac{g}{4}\left[\lambda_1
\frac{k_x+q_x/2-i(k_y+q_y/2+\mu_0B/\alpha)^2}{|\mathbf{k}+\mathbf{q}/2+\mu_0B\hat{y}/\alpha|}+
\lambda_2
\frac{k_x-q_x/2-i(k_y-q_y/2-\mu_0B/\alpha)^2}{|\mathbf{k}-\mathbf{q}/2-\mu_0B\hat{y}/\alpha|}
\right]\\&\times
\left[\lambda_3
\frac{k_x+q_x/2+i(k_y+q_y/2+\mu_0B/\alpha)^2}{|\mathbf{k}+\mathbf{q}/2+\mu_0B\hat{y}/\alpha|}+
\lambda_4
\frac{k_x-q_x/2+i(k_y-q_y/2-\mu_0B/\alpha)^2}{|\mathbf{k}-\mathbf{q}/2-\mu_0B\hat{y}/\alpha|}
\right].
\end{align}  
\end{subequations}

\begin{figure}[ttt]
\begin{center}
\includegraphics[width=3.5in]{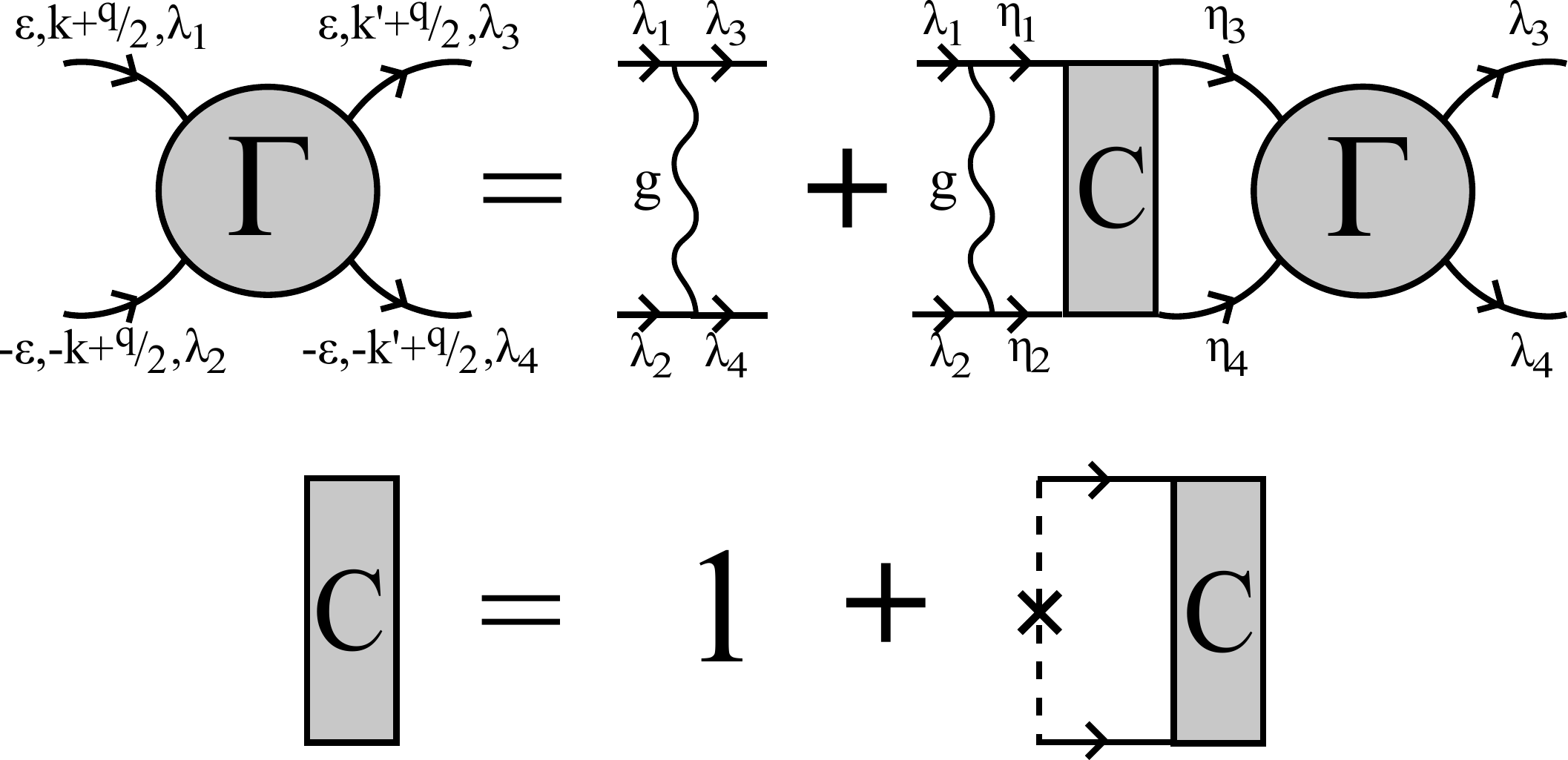}
\end{center}
\vspace{-0.1in}
\caption{Diagrammatic representation of Cooper pair propagator (top line) and Cooperon ladder (bottom line). Straight lines with arrows are electron Green's functions, wiggly lines represent attractive phonon mediated interactions, dashed lines with $\times$ represent disorder scattering, and Greek letters label the $\pm$ Rashba bands.}
\label{fig:CooperVertex}
\vspace{0.1in}
\end{figure}

We are now equipped to study the vertex equation for the interaction in the Cooper channel. This equation takes the form:
\begin{align}\label{eq:Vertex_equation}
\Gamma_{\lambda_1,\lambda_2,\lambda_3,\lambda_4}(\mathbf{k},\mathbf{k}',\mathbf{q})=g_{\lambda_1,\lambda_2,\lambda_3,\lambda_4}(\mathbf{k},\mathbf{k}',\mathbf{q})+\sum_{\mathbf{k}_1\e_1}g_{\lambda_1,\lambda_2,\eta_1\eta_2}(\mathbf{k},\mathbf{k}_1,\mathbf{q})
\mathcal{K}_{\eta_1,\eta_2,\eta_3,\eta_4}(\mathbf{k}_1,\mathbf{q})
\Gamma_{\eta_3,\eta_4,\lambda_3,\lambda_4}(\mathbf{k}_1,\mathbf{k}',\mathbf{q}).
\end{align}  
Here $\mathcal{K}$ denotes the two particle Green's function in the particle-particle channel. In the presence of impurities, $\mathcal{K}$ must include scattering by the disorder potential. Therefore, in the ladder approximation, the two particle Green's function is a product of two Green's function connected through the Cooperon, $C_{\eta_1,\eta_2,\eta_3,\eta_4}(\mathbf{q},\e_n)$:   
\begin{align}\label{eq:Cooperon}
\mathcal{K}_{\eta_1,\eta_2,\eta_3,\eta_4}(\mathbf{k},\e_n,\mathbf{q})=G_{\eta_1}(\mathbf{q}/2+\mathbf{k},\e_n)G_{\eta_2}(\mathbf{q}/2-\mathbf{k},-\e_n)C_{\eta_1,\eta_2,\eta_3,\eta_4}(\mathbf{q},\e_n),
\end{align}
where $\eta_{1,2}$ are not implicitly summed.
The single electron Green's function can be written in terms of the eigenenergies of the two branches given in Eq.~\ref{eq:Chiral_Basis_Energy}, 
\begin{equation}G_{\lambda}(\mathbf{k},\e_n)=\[i\e_n-E_{\lambda}(\mathbf{k})+\mu+\frac{i}{2\tau}\text{sign}(\e_n)\]^{-1} .
\label{eq:DisorderGreensFn}
\end{equation} 
The quasiparticle life time is a function of the density of states of both bands,  $1/\tau=V^2\pi(\nu_{+}+\nu_{-})$ with  $\nu_{\lambda} = m(1-\lambda{\alpha}/{v_F})/2\pi$. The vertex equation is illustrated diagrammatically in Fig.~\ref{fig:CooperVertex}. The critical exchange field is determined by the largest field for which there is still an instability towards pairing, i.e. for which the effective interaction in the Cooper channel still diverges. For $|\mathbf{q}|,\mu_0B/\alpha\ll{k_F}$, we can neglect the dependence of the bare vertex  (Eq.~\ref{eq:Chiral_Basis_g}) on $\mathbf{q}$ and $B$. Then, the gap equation takes the following simple form: 
\begin{align}\label{eq:Vertex_equation2}
\Gamma_{\lambda_1,\lambda_2,\lambda_3,\lambda_4}(\mathbf{k},\mathbf{k}',\mathbf{q})=g_{\lambda_1,\lambda_2,\lambda_3,\lambda_4}(\mathbf{k},\mathbf{k}',\mathbf{q})\left[1-g\sum_{\mathbf{k}_1,\e_1}\mathcal{K}_{\eta_1,\eta_1,\eta_2,\eta_2}(\mathbf{k}_1,\mathbf{q})
\right]^{-1}.
\end{align}  
Note that although the vertex only includes contributions where the two incoming (outgoing) electrons are from the same Rashba branch, the Cooperon contains sections in which the two electrons occupy different branches.  We will not show here the general solution for the vertex equation, as it is rather cumbersome. Instead, we will present the expression for the vertex in four different regimes and extract the corresponding values of the critical exchange field at $T=0$. We wish to point out that since the exchange field enters the energy by shifting the momentum $k_y$ inside the spin-orbit term (see Eq.~\ref{eq:Chiral_Basis_Energy}), the FFLO state have finite momentum along the $y$ direction, $\mathbf{q}=q\hat{y}$.
 
\section{Pairing Interaction Vertex Without Disorder}

We first analyze the Cooper vertex in the absence of disorder ($\tau^{-1}=0$). In this limit, the Cooperon is equal to unity, and the vertex described in Eq.~\ref{eq:Vertex_equation2} diverges when:
\begin{align}\label{eq:Gap_equation_clean}
1-g\sum_{\mathbf{k}_1,\e_n}[G_{+}(\mathbf{q}/2+\mathbf{k}_1,\e_n)G_{+}(\mathbf{q}/2-\mathbf{k}_1,-\e_n)+G_{-}(\mathbf{q}/2+\mathbf{k}_1,\e_n)G_{-}(\mathbf{q}/2-\mathbf{k}_1,-\e_n)]=0
\end{align}
One can see that the onset of pairing occurs simultaneously in both bands.  To simplify the integration over the momentum and frequency, we expand the single particle energies in powers of $q/k_F$ and $B/\alpha{k_F}$: $G_{\lambda}(\mathbf{q}/2+\mathbf{k},\e_n)\approx[i\e_n-\frac{k^2}{2m}+\lambda\alpha|\mathbf{k}|+\mu+(v_Fq-\lambda\mu_0B)]^{-1}$. To linear order, the interaction vertex becomes:
\begin{align}\label{eq:Gap_equation_clean2}
\Gamma^{-1}\approx 1-g\left[\nu_{+}\ln\frac{\omega_D}{|v_F{q}/{2}+\mu_0B|}+\nu_{-}\ln\frac{\omega_D}{|v_F{q}/{2}-\mu_0B|}\right]
\end{align}
For $q=0$ and in the absence of the Zeeman field $B$ the renormalized interaction diverges logarithmically.  For finite $q$ and $B$, the odd power terms in the expansion of $\e_{{\bf q}/2\pm{\bf k}}^\lambda$ have opposite signs,  and hence, they cut off the logarithmic divergence.  

Because the density of states is larger in the $-$ band, the interaction vertex is most divergent for: 
\begin{equation} 
q_{\text{opt}}^{\text{(clean)}} =  \frac{2\mu_0B}{v_F} \label{eq:QoptClean}
\end{equation}
To linear order in the magnetic field and the pair momentum, this choice of $q$ cancels the effect of the magnetic field on pairs in the $-$ band.   With this choice, the logarithmic divergence in the $-$ branch is cut off only by terms of order $\mathcal{O}\({B^3}/{\Delta_{so}^2}\)$.  The critical field $B_c$ is the field beyond which the interaction vertex is finite:
\begin{align}\label{eq:Gap_equation_clean3}
\ln{\Delta_0}\approx \nu_{+}\ln(\mu_0B_c)+\nu_{-}\ln\frac{(\mu_0B_c)^3}{\Delta_{so}^2},
\end{align} 
where $\Delta_0=\omega_D\exp\(\frac{-1}{(\nu_{+}+\nu_{-})g}\)=\omega_D\exp\(\frac{-\pi}{mg}\)$ is the gap in the absence of the exchange field, and  $\Delta_{so}=2\alpha k_F$.  From Eq.~\ref{eq:Gap_equation_clean3}, we identify:
 \begin{align}\label{eq:Bc_clean}
\mu_0B_c\approx\Delta_0\left(\frac{\Delta_{so}}{\Delta_0}\right)^{\frac{1+\alpha/v_F}{2+\alpha/v_F}}.
\end{align} 

\subsection{Anisotropic Bands}
We now re-examine the possibility of finite-momentum pairing in systems with anisotropic band dispersions.  This is relevant, for example, in the oxide interface system where we expect superconductivity to arise predominantly in bands descending from $d_{xz}$ and $d_{yz}$ orbitals\cite{CompanionPaper}, which have anisotropic hopping amplitudes in the $xy$ plane.  We model the band-anisotropy by introducing different effective masses for the $x$ and $y$ directions, replacing the kinetic energy term by:
\begin{align}\label{eq:AnisotropicKE} \frac{k^2}{2m}\rightarrow \frac{k_x^2}{2m_x}+\frac{k_y^2}{2m_y} = \frac{k^2}{2\tilde{m}}\(1-\epsilon \cos 2\phi_k\)  \end{align}
where $\tilde{m} = \frac{2m_xm_y}{m_x+m_y}$, $\epsilon = \frac{m_x-m_y}{2(m_x+m_y)} \in (-1,1)$ is the eccentricity of the kinetic term, and $\phi_k$ is the angle between the momentum and the $x$-axis.

The effect of anisotropic bands is easiest to analyze in the limits $\Delta_{so}\gg \mu$, and $\Delta_{so}\ll \mu$. For the case of $\Delta_{so}\gg \mu$, the problem is essentially identical to the isotropic case described above, once we replace $m\rightarrow m_y$ in Eq. \ref{eq:QoptClean},\ref{eq:Bc_clean}.  In this limit, the optimal pairing wave-vector becomes $q_{\text{opt}} = \frac{\mu_0B}{v_{F,y}}$, and the critical field is related to that of the isotropic band case by: $\mu_0B_c^{\text{anis.}} \approx \(\frac{m_y}{m}\)^{2/3}\mu_0B_c^{\text{iso.}}$.  In this limit, the critical field is either enhanced (suppressed) relative to the isotropic case if the band has larger (smaller) band-mass in the direction perpendicular to the exchange field, $\mathbf{B}$.

In the opposite limit, $\Delta_{so}\ll \mu$,  one can find a closed form expression for the critical field $B_c$ for weak anisotropy ($\epsilon\ll 1$):
$B_c^{\text{anis}}\approx \(1+\frac{\pi\epsilon^2}{16}\)B_c^{\text{iso}}$.  In this limit, the critical field is enhanced by the anisotropy independent of the direction of the anisotropy.  This can be understood simply by noting that the anisotropic dispersion leads to an elongated Fermi-surface, leading to an enhancement of nesting in the Cooper channel for pairs forming along the elongated direction.

Generically, we find that the anisotropy can shift the quantitative values of the pairing vector $q_\text{opt}$ and the critical field $B_c$, but do not alter the parametric enhancement of the critical field due to spin-orbit.  We expect that band-anisotropy will be similarly unimportant in the presence of disorder.

\section{Pairing Interaction Vertex with Disorder}

Having described pairing in the absence of disorder, we now examine how the  analysis is modified in the presence of disorder.  Scattering by impurities broadens the bands, giving the quasiparticles finite (elastic) life-times $\tau$. The finite life-time contains contributions from inter- as well as intra-band scattering.  From the general theory of superconductivity we know that disorder does not affect the superconducting state as long as the Cooper pairs remain correlated after each scattering event. In other words, if the probability of an electron in state $\mathbf{k}$ to be scattered into state $\mathbf{k}'$ is equal to the probability of its time reversed partner  to be scattered into the partner of $\mathbf{k}'$.   Quantitatively, these correlation are described by the Cooperon (see Eq.~\ref{eq:Cooperon} and subsequent discussion). Fully correlated pair scattering manifests  itself in a massless Cooperon, i.e., the Cooperon, having a diffusive pole, diverges at zero frequency and momentum.  As long as $\Delta_{so}\tau< 1$, the pair scattering is dominated by processes where both electrons either remain in the same Rashba band or transfer to the other band, and, in principle, the Cooperon could have a diffusive pole. In this scenario, the superconducting state would not be sensitive to disorder. However, as we show below, the magnetic field reduces the pair correlation after certain scattering events, and  the Cooperon is massive. This effect is most relevant  for weak disorder.

An additional effect of  disorder is related to spin precession. As a consequence of the spin orbit coupling, in the absence of magnetic field,  the spin of an electron with momentum $\mathbf{k}$ precesses with frequency $\Delta_{so}$ around the vector $\hat{k}$.  Since time reversal symmetry is preserved,  electrons with $\pm\mathbf{k}$  can pair to create a superconducting state.  When a magnetic field is applied and time reversal symmetry is broken, the magnetic field  changes the magnitude and direction of the spin precession (we consider here only the limit $\mu_0B\ll\Delta_{so}$ where the spin precession due to the spin-orbit coupling is weakly modified  by the magnetic field). Correspondingly, the electrons with momenta $\pm\mathbf{k}$ are no longer related by time reversal (the $\pm\mathbf{k}$ pairs dephase), and the magnetic field serves as pair breaking mechanism.   Note that changing the pair momentum $\pm\mathbf{k}\rightarrow\pm\mathbf{k}+\mathbf{q}/2$ not only effect the spin precession, it also changes the kinetic energy of the pair. Superconductivity is obtained when these two dephasing mechanisms are minimized together. As a consequence of scattering by impurities the spin precession relaxes. Each time an electron is scattered into a different momentum state, it experiences a different torque which  leads to relaxation of the precession.  For disorder weaker than the spin orbit coupling, the large  (time reversal preserving) component of the spin precession  is almost not affected by the impurities. Nevertheless,  the additional (breaking time reversal) precession of the spin due to the magnetic field can be suppressed by disorder. The latter, being relevant in the regime  $\mu_0B<\tau^{-1}$, actually helps the superconducting state to survive at  high magnetic fields, as we show below. 

Taking all elements into account, we identify three distinct regimes: 1) weak disorder $\Delta_0\tau> 1$, 2) moderate disorder $\Delta_0<\tau^{-1}<\Delta_{so}$, and 3) strong disorder $\Delta_{so}\tau< 1$, and analyze each in turn.  The first two regimes were studied in \cite{DimitrovaSupp} in the limit of weak spin-orbit coupling.  Here we extend the analysis to all regimes to treat arbitrary strengths of spin-orbit and disorder.

\subsection{Weak Disorder ($\tau\Delta_0>1$)}
 
For weak disorder,  $\tau^{-1}\ll \tau_B^{-1}\approx\mu_0B_c$, we can assume that just like in the clean case, $|v_Fq/2-\mu_0B|<\tau^{-1}$, while $|v_Fq/2+\mu_0B|>\tau^{-1}$ (as we will see this assumption is consistent with our solution). Then, the pair coherence is strongly suppressed  after scattering from the $-$ to the $+$ band and  the Cooperon is massive (with mass of the order $\tau^{-1}$).  Consequently the Cooperon correction can be neglected, and   Eq.~\ref{eq:Gap_equation_clean} is only modified by the quasiparticle life-time entering the single particle Green's functions  (see Eq. \ref{eq:DisorderGreensFn}). Integrating over the frequency and momentum in Eq. \ref{eq:Gap_equation_clean}, we find that $\Gamma$ diverges when:
\begin{align}\label{eq:Gap_equation_weak_disorder}
-\frac{1}{g}=\nu_{-}\ln\frac{\omega_D}{1/2\tau+\sqrt{1/4\tau^2+(v_Fq-2\mu_0B)^2)}}+\nu_{+}\ln\frac{\omega_D}{1/2\tau+\sqrt{1/4\tau^2+(v_Fq+2\mu_0B)^2)}}.
\end{align}  
The optimal pairing momentum is identical to the one obtained in the clean limit $\mathbf{q}_{\text{opt}}=2\mu{B}\hat{y}/v_F$. The critical field, on the other hand, is strongly suppressed, $\mu_0B_c\approx\Delta_0(\Delta_0\tau)^{\frac{1+\alpha/v_F}{1-\alpha/v_F}}$. This expression for the critical field is valid as long as $(\mu_0B_c)^3/\Delta_{so}^2<1/\tau<B_c$. The critical exchange field decreases as the disorder increases, and as the system  approaches the limit $1/\tau\approx\Delta_0$  the critical field gets closer to the Pauli limit, $\mu_0B_c\approx\Delta_0$. 

In this weak disorder limit, the fact that the quasiparticle life-time enters the gap equation through the single particle Green's function for the $-$ band, while the Cooperon is massive tells us that some scattering events suppress pair coherence. More precisely, as long as the two electrons forming a Cooper pair scatter between states in the $-$ band, they remain correlated. However, the probability of an electron to be scattered from one band to  the other is not negligible.  Scattering of a Cooper pair from the $-$ to the $+$ band (where the magnetic field serves as a strong pair breaking mechanism), results in decoherence of the pair. 
[One can check that if scattering is limited to a single Rashba band, the FFLO state is immune to weak disorder.]

\subsection{Moderate Disorder ($\Delta_0<\tau^{-1}<\Delta_{so}$) -- ``Diffusive Regime"}

In the previous section we saw that weak disorder strongly suppresses the critical field as $\tau^{-1}\rightarrow \Delta_0$.  One might expect that stronger disorder could only further weaken the FFLO state. Surprisingly the finite momentum pairing re-emerges for moderate disorder $\Delta_0<\tau^{-1}<\Delta_{so}$, and the critical field increases beyond $\Delta_0$. This regime has been studied in Ref.~\cite{DimitrovaSupp} for the weak spin orbit limit, $\alpha\ll v_F$. The key element is that $\tau_B>\tau$. Then, the component of the spin precession caused by the magnetic field is diffusive, and hence, strongly suppressed. Similar to the effect of scattering on momentum, the frequent  changes in torque after each scattering event  leads to diffusion of the vector of spin precession. As a consequence of the diffusion of both spin precession and momentum, the mechanism that previously reduced significantly the pair coherence gets weaker with increasing disorder. Correspondingly,  the magnetic field enters the Cooperon as $\tau_{B}^{-2}\tau\ll\tau^{-1}$, i.e. in the same way as the pair momentum $\mathbf{q}$.  Therefore, the effect of $B$ is smaller than in  the weak disorder limit, and the system can tolerate higher magnetic  fields.  This is the main difference between the  FFLO state in systems with and without spin-orbit coupling. In the absence of spin-orbit coupling, the magnetic field appears in the Cooperon as $1/\tau_B$.

Incorporating the Cooperon in the vertex equation, the instability point is determined by: 
\begin{align}\label{eq:Gap_equation_mod_disorder}
\frac{1}{g}=\sum_{\e_n}\frac{\overline{G_{+}G_{+}(0)}+\overline{G_{-}G_{-}(0)}}{1-V^2\overline{G_{+}G_{+}(0)}/2-V^2\overline{G_{-}G_{-}(0)}/2+V^4[\overline{G_{+}G_{+}(1)}-\overline{G_{-}G_{-}(1)}]^2/2}.
\end{align}  
In the above equation and below we use the notation $\overline{G_{\lambda}G_{\lambda'}(m)}$ for the m-th angular-harmonic of the two Green's function averaged over the momentum: 
\begin{align}\label{eq:Avr_GFs}
\overline{G_{\lambda}G_{\lambda'}(m)}=\int\frac{d\mathbf{k}}{(2\pi)^d}e^{-im\theta}G_{\lambda}(\mathbf{q}/2+\mathbf{k}_1,\e_n)G_{\lambda'}(\mathbf{q}/2-\mathbf{k}_1,-\e_n).
\end{align}   
Performing the frequency integration in  Eq.~\ref{eq:Gap_equation_mod_disorder}, we get: 
\begin{align}\label{eq:Gap_equation_mod_disorder2}
\ln{\Delta_0}=\ln\left[\frac{\nu_{+}}{4\nu_0}(v_Fq+2\mu_0B)^2\tau+\frac{\nu_{-}}{4\nu_0}(v_Fq-2\mu_0B)^2\tau+\frac{1}{2}(\alpha{q}-2\mu_0B)^2\tau\right].
\end{align}  
The instability still occurs at finite momentum but the pair momentum in the diffusive regime is  different  than the one obtained in the clean case:
\begin{equation}
 q_{\text{opt}}^{(\text{diff})}=\cfrac{4\alpha \mu_0B}{\alpha^2+v_F^2}
 \end{equation}
In this regime, the optimal $q$ is not simply given by optimizing one Rashba branch at the expense of the other.  Rather, it is chosen to compromise between both branches in order to minimize the frequency in which the Cooperon diverges.  If both branches had the same densities of states, pairing would occur at zero momentum.  In the limit $\alpha\ll v_F$, we recover the result of Ref.~\cite{DimitrovaSupp}. 

The  critical field corresponding to $ q_{\text{opt}}^{(\text{diff})}$ is $B_c\approx\Delta_0(\Delta_0\tau)^{-1/2}$, which grows with increasing disorder almost recovering its value in the clean limit as $\tau^{-1}\rightarrow \Delta_{so}$.

\subsection{Strong Disorder ($\tau^{-1}\gg \Delta_{so}$)}

The revival of the FFLO state in the limit of moderate disorder can be attributed to the diffusive behavior of both the electron momentum and spin. For even stronger impurity scattering $1/\tau\gg\Delta_{so}$, the precession picture described previously breaks down.   More concretely, while the momentum relaxation remains to be  $\tau^{-1}$, the spin relaxation becomes smaller due to motional narrowing $\tilde\tau_{so}^{-1}\approx\Delta_{so}^2\tau $ (the Dyakonov-Perel mechanism~\cite{DyakonovSupp}).  In the extreme limit $\Delta_{so}\tau\rightarrow 0$, the  direction of the spins decouples from the momentum and the spin--orbit coupling is completely ineffective.  We then expect the critical field to approach the Pauli limit.  

These expectations are confirmed in calculating the vertex of the attractive interaction.  Indeed, in the strong disorder limit, the Cooperon loses its diffusive behavior, and the gap equation is,
\begin{align}\label{eq:Gap_equation_strong_disorder}
\Gamma^{-1}=1-g\sum_{\e_n}&[\overline{G_{+}G_{+}(0)}+\overline{G_{-}G_{-}(0)}]
\left\{\left[1+\frac{A-B}{2\sqrt{(A-B)^2-C^2}}\right]\left[A+B-\sqrt{(A-B)^2-C^2}\right]^{-1}\right.\\\nonumber
&\left.
+\left[1-\frac{A-B}{2\sqrt{(A-B)^2-C^2}}\right]\left[A-B-\sqrt{(A-B)^2-C^2}\right]^{-1}\right\},
\end{align}  
where
\begin{align}\label{eq:Gap_equation_strong_disorder2}
&A=1-\frac{V^2}{2}\left[\overline{G_{+}G_{+}(0)}+\overline{G_{-}G_{-}(0)}\right]+\left(\frac{\alpha{q}+2\mu_0B}{2\Delta_{so}}\right)^2;\\\nonumber
&B=1-\frac{V^2}{4}\left[\overline{G_{+}G_{+}(0)}+\overline{G_{-}G_{-}(0)}+\overline{G_{-}G_{+}(0)}+\overline{G_{+}G_{-}(0)}\right]+\left(\frac{\alpha{q}+2\mu_0B}{2\Delta_{so}}\right)^2;\\\nonumber
&C=\frac{V^2}{2}\left[\overline{G_{+}G_{+}(1)}-\overline{G_{-}G_{-}(1)}\right].
\end{align} 
Integrating Eq.~\ref {eq:Gap_equation_strong_disorder} over the frequency, the condition for the onset of pairing becomes:
\begin{align}\label{eq:Gap_equation_strong_disorder3}
\ln{\Delta_0}&=\frac{1}{2}
\left[1+\frac{\Delta_{so}^2\tau}{\sqrt{\Delta_{so}^4\tau^2-(\alpha{q}-2\mu_0B)^2/4}}\right]\ln\left[\Delta_{so}^2\tau+\frac{(\alpha{q}+2\mu_0B)^2}{\Delta_{so}^2\tau}-\sqrt{\Delta_{so}^4\tau^2-(\alpha{q}-2\mu_0B)^2/4}\right]\\\nonumber
&+\frac{1}{2}
\left[1-\frac{\Delta_{so}^2\tau}{\sqrt{\Delta_{so}^4\tau^2-(\alpha{q}-2\mu_0B)^2/4}}\right]\ln\left[\Delta_{so}^2\tau+\frac{(\alpha{q}+2\mu_0B)^2}{\Delta_{so}^2\tau}+\sqrt{\Delta_{so}^4\tau^2-(\alpha{q}-2\mu_0B)^2/4}\right]
\end{align}  
In the above expression we have dropped terms of the order $(v_Fq\tau)^2,(B\tau)^2\ll1$. For $q=0$ our expression coincides with the one derived in Ref.~\cite{KlemmSupp}.  We can see that all of the natural choices for pairing momentum: $q=0$ and $q=\pm{2\mu_0B}/\alpha$, give rise to critical fields of the same order:
$B_c\approx\Delta_0\sqrt{\Delta_{so}^2\tau/\Delta_0}\approx\Delta_0(\Delta_{0}\tilde{\tau}_{so})^{-1/2}$ which eventually decays to the Pauli limit as disorder becomes infinitely strong. To determine which of the three possible values of $q$ is the optimal pair momentum, one has to examine more carefully the numerical coefficients in the corresponding expressions for $B_c$.  Nevertheless, as $1/\tau$, becomes large enough, we expect to have a first order phase transition from the FFLO state into a superconducting state with $q=0$.

\section{Discussion}
To summarize, we have presented the derivation of the critical field as a function of disorder for four different regimes. We have shown that scattering by impurities strongly  affects the superconducting state with finite pairing momentum. While weak disorder significantly reduces    the critical exchange field, the diffusive behavior obtained when $1/\tau>\Delta_{0}$ helps to increase the field almost up  to its value in the clean limit. However, crossing to the regime of strong disorder,  $1/\tau>\Delta_{so}$, the direction of the spin is no longer correlated with the momentum and there is no difference between a system with and without spin-orbit coupling. Therefore, the FFLO state disappears at strong disorder, and the critical field approaches the Pauli limit.


\begin{thebibliography}{99}
\bibitem{Ohtomo04}
A. Ohtomo and H.Y. Hwang, Nature {\bf 427}, 423 (2004).

\bibitem{Reyren07}
N. Reyren {\em et al}., Science {\bf 317}, 1196 (2007).

\bibitem{Brinkman07}
A. Brinkman {\em et al}., Nat. Materials, 6, 493 (2007).

\bibitem{Dikin}
D.A. Dikin {\em et al}., Phys. Rev. Lett. {\bf 107}, 056802 (2011).

\bibitem{Li}
L. Li, C. Richter, J. Mannhart and R.C. Ashoori, Nat. Phys. {\bf 7}, 762 (2011).

\bibitem{Bert11}
J.~A.~Bert, {\em et al}.,  Nat. Phys. {\bf 7}, 767 (2011).

\bibitem{Ariando11}
Ariando, {\em et al}.,   Nat Commun {\bf 2}, 188 (2011).

\bibitem{Okamoto06}
S. Okamoto, A.J. Millis and N.A. Spaldin, Phys. Rev. Lett {\bf 97}, 056802 (2006). 

\bibitem{Pentcheva07/08}
R. Pentcheva, and W.E. Pickett, Phys. Rev. Lett. {\bf 99}, 016802 (2007); Phys. Rev. B {\bf 78}, 205106 (2008).

\bibitem{Joshua11}  A.~Joshua, S~ Pecker, J.~ Ruhman, E.~Altman and S.~Ilani ArXiv: 1110.2184.

\bibitem{Popovic08}
Z.S. Popovic, S. Satpathy and R.M. Martin, Phys. Rev. Lett. {\bf 101}, 256801 (2008).

\bibitem{Caviglia10}
A.D. Caviglia {\em et al}., Phys. Rev. Lett. {\bf 104},  126803 (2010).

\bibitem{Fulde64}
F. Fulde and R.A. Ferrell, Phys. Rev. {\bf 135}, A550 (1964).

\bibitem{Larkin65}
A.I. Larkin and Y.N. Ovchinikov, Sov. Phys. JETP {\bf 20}, 762 (1965).

\bibitem{Barzykin02}
V. Barzykin and L.P. Gorkov, Phys. Rev. Lett. {\bf 89}, 227002 (2002).

\bibitem{Dimitrova}
O. Dimitrova and M.V. Feigel'man, Phys. Rev. B {\bf 76} 014522 (2007).

\bibitem{BenShalom10}
M. BenShalom. A. Ron, A. Palevski and Y. Dagan, Phys. Rev. Lett. {\bf 105}, 206401 (2010). 

\bibitem{Bell09} 
C. Bell {\em et al}., Phys. Rev. Lett. {\bf 103}, 226802 (2009).

\bibitem{Nagano09}
See M. Nagano, A. Kodama, T. Shishidou and T. Oguchi, J. Phys: Cond. Matt. {\bf 21}, 064239 (2009).

\bibitem{Dietl97}
T. Dietl, A. Haury and Y.M. d'Aubigne, Phys. Rev. B {\bf 55}, R3347 (1997).

\bibitem{Zener51}
C. Zener, Phys. Rev. {\bf 81}, 440 (1951).

\bibitem{Jungwirth06}
For a review, see T. Jungwirth {\em et al}., Rev. Mod. Phys. {\bf 78}, 809 (2006).

\bibitem{Yip02}
S.K. Yip, Phys. Rev. B {\bf 65}, 144508 (2002).

\bibitem{GSCurrentEndNote}
Despite the presence of a single, finite $\mathbf{q}$, the ground-state carries zero-current\cite{Dimitrova}.  

\bibitem{AnisotropicDispersionSupplement}
In reality, the $d_{xz}$ and $d_{yz}$ bands will have anisotropic dispersions. However, as demonstrated in the supplementary material at EPAPS:$\#$, band anisotropy does not introduce any important changes to our model.  Therefore  we choose to work with a simpler isotropic dispersion.  

\bibitem{SignAlphaEndnote}
Here and throughout we have assumed $\alpha>0$, for $\alpha<0$ one should replace $\mathbf{q}\rightarrow -\mathbf{q}$.

\bibitem{Dyakonov}
M.~I,~D'yakonov and V.~I.~Perel, Zh. Eksp. Teor. Fiz. 60, 1954 (1971)
[Sov. Phys. JETP 30, 1053 (1971)].

\bibitem{Klemm}
R.~A.~Klemm, A. Luther and M.~R.~Beasley, Phys. Rev. B {\bf 12}, 877 (1975). 

\end{thebibliography}

\begin{thebibliography}{99}
\bibitem{CompanionPaperSupp}
In preparation.

\bibitem{DimitrovaSupp} O.~Dimitrova and M.~V.~FeigelÕman, Phys. Rev. B {\bf 76},  014522 (2007).    


\bibitem{DyakonovSupp}
M.~I,~D'yakonov and V.~I.~Perel, Zh. Eksp. Teor. Fiz. 60, 1954 (1971)
[Sov. Phys. JETP 30, 1053 (1971)].

\bibitem{KlemmSupp}
R.~A.~Klemm, A. Luther and M.~R.~Beasley, Phys. Rev. B {\bf 12}, 877 (1975). 

\end{thebibliography}
\end{document}